\documentclass[11pt,a4paper]{article}
\usepackage[utf8]{inputenc}
\usepackage[english]{babel}

\usepackage[left=2.5cm,right=2.5cm,top=2.5cm,bottom=2.5cm]{geometry}
\usepackage{url}

\usepackage[electronic]{ifsym}
\usepackage{indentfirst,latexsym,bm,bbm}
\usepackage{amsmath,amssymb,amsfonts,amsthm,amscd,amssymb}
\usepackage{pifont,wasysym}
\usepackage[mathscr]{euscript}

\usepackage[rgb]{xcolor}
\usepackage{graphicx}
\usepackage{subfigure}
\usepackage{algorithmic}
\usepackage[noend, ruled, linesnumbered]{algorithm2e}

\DeclareMathOperator{\dif}{d}

\renewcommand{\vec}[1]{\bm{#1}}

\newcommand{\mat}[1]{\bm{#1}}

\newcommand{\mj} {\mathrm{j}}

\newcommand{\me} {\mathrm{e}}
\newcommand{\fracpde}[2]{\frac{\partial {#1}}{\partial {#2}}}
\newcommand{\fracode}[2]{\frac{\dif {#1}}{\dif {#2}}}
\newcommand{\fracpdemix}[3]{\frac{\partial^2 {#1}}{\partial {#2} \partial {#3}}}

\newcommand{\secode}[2]{\frac{\dif^2 {#1}}{\dif {#2}^2}}
\newcommand{\set}[1]{\left\{ #1 \right\}}
\newcommand{\abs}[1]{\left| #1 \right|}
\newcommand{\norm}[1]{\left\lVert #1 \right\rVert}

\newcommand{\trsp}[1]{{#1}^\textsf{T}}
\newcommand{\inv}[1]{#1^{-1}}

\newcommand{\vecasym}[1]{[#1]_\times}   % antisymmetric matrix from a vector
\newcommand{\GL}[2]{\mathrm{GL}(#1,\mathbb{#2})}
\newcommand{\ES}[3]{\mathbb{#1}^{{#2}\times {#3}}}
\DeclareMathAlphabet{\mathsfsl}{OT1}{cmss}{m}{sl}
\newcommand{\quat}[1]{\mathbbm{#1}}       % quaternions
\newtheorem{thm}{Theorem}

\newtheorem{lem}[thm]{Lemma}

\begin{document}

\title{\textbf{Symplectic Geometric Algorithm for Quaternion Kinematical Differential Equation}}

\author{
Hong-Yan Zhang\footnote{Corresponding author, Dr. Hong-Yan Zhang, Email: hongyan.rd@foxmail.com},
Lu-Sha Zhou,
Zi-Hao Wang,
Long Ma
and Yi-Fan Niu \\ ~~\\
Civil Aviation University of China, Tianjin, P. R. China, 300300
}
\date{}

\maketitle
\begin{abstract}
Solving quaternion kinematical differential equations is one of the most significant problems in the automation, navigation, aerospace and aeronautics literatures.  Most existing approaches for this problem neither preserve the norm of quaternions nor avoid errors accumulated in the sense of long term time.  We present symplectic geometric algorithms to deal with the quaternion kinematical differential equation by modeling its time-invariant and time-varying versions with Hamiltonian systems by adopting a three-step strategy.  Firstly, a generalized Euler's formula for the autonomous quaternion kinematical differential equation are proved and used to construct symplectic single-step transition operators via the centered implicit
Euler scheme for autonomous Hamiltonian system. Secondly, the symplecitiy, orthogonality and invertibility of the symplectic transition operators are proved rigorously. Finally, the main results obtained are generalized to design symplectic geometric algorithm for the time-varying quaternion kinematical differential equation which is a non-autonomous and nonlinear Hamiltonian system essentially. Our novel algorithms have simple algorithmic structures and low time complexity of computation, which are easy to be implemented with real-time techniques.  The correctness and efficiencies of the proposed algorithms are verified and validated via numerical simulations.

\end{abstract}

\section{Introduction}

Quaternions, invented by the Irish mathematician W. R. Hamiltonian in 1843, have been extensively utilized
in physics \cite{Robinson1958, Lu2006}, aerospace and aeronautical technologies  \cite{Wie1985,Wie1989,Wie1995,Wie2002,Kuipers2002,Friedland1978,Kim2004,Rogers2007,Zhong2012,Zamani2013},
robotics and automation \cite{Yuan1988,Funda1990,Chou1992,Fjellstad1994,Cheng201615}, human motion capture \cite{Alexiadis2014}, computer graphics and games \cite{Dunn2002,Eberly2007,Vince2011}, molecular dynamics \cite{Miller2002}, and flight simulation \cite{Cooke1994,Allerton2009,Diston2009,JSBSim2014}. Quaternions have no inherent geometrical
singularity as Euler angles when parameterizing the 3-dimensional special orthogonal group manifold $\mathrm{SO}(3,\mathbb{R})$  with local coordinates and
they are useful for real-time computation since only simple multiplications and additions are needed
instead of trigonometric relations. Almost all of the  researches available about the applications of quaternions
focus on these two merits and the fundamental \textit{quaternion kinematical differential equation} (QKDE) \cite{Robinson1958,Friedland1978,Wie1985}. In \cite{Robinson1958} (see  page-21, Eq.77.), Robinson presented the following QKDE
\begin{equation} \label{eq-QKDE}
\left\{
\begin{split}
&\fracode{\quat{q}}{t} = \frac{1}{2} \mat{A}(\vec{\omega}(t))\cdot \quat{q}, \quad t> t_0 \\
&\quat{q}(t_0) = \quat{q}_0
\end{split}
\right.
\end{equation}
where $t_0$ is the initial time,
$\vec{\omega} =\trsp{[\omega_1(t), \omega_2(t), \omega_3(t)]}$
is the angular velocity vector,
$\quat{q}  = \trsp{[e_0, e_1, e_2, e_3]}$
is the matrix representation of the quaternion $\quat{q}$ with scalar part $e_0$ as well as vector part $\trsp{[e_1, e_2, e_3]}$,
and
\begin{equation} \label{eq-A}
\begin{split}
\mat{A}
&= \mat{A}(\vec{\omega}(t))=\begin{bmatrix} 0 & -\trsp{\vec{\omega}} \\ \vec{\omega} & -\vecasym{\vec{\omega}} \end{bmatrix}
=-\trsp{\mat{A}}, \\
\vecasym{\vec{\omega}} &= \begin{bmatrix} 0 & -\omega_3 &\omega_2 \\ \omega_3 & 0 & -\omega_1 \\ -\omega_2 & \omega_1 & 0 \end{bmatrix}
= -\trsp{(\vecasym{\vec{\omega}})}.
\end{split}
\end{equation}
Formally, Eq.(\ref{eq-QKDE}) is a linear ordinary differential equation (ODE) and its numerical solution should be easily determined in practical engineering applications. However, as Wie and Barbar \cite{Wie1985} pointed out, the coefficients $\omega_1, \omega_2, \omega_3$, or equivalently the angular rate vector $\vec{\omega}=\trsp{[\omega_1, \omega_2, \omega_3]}$,  are time-varying and  the matrix $\mat{A}$ has the repeated eigenvalues
$$
\pm \mj \norm{\vec{\omega}} = \pm \mj \sqrt{\omega_1^2 + \omega_2^2 + \omega_3^2},$$
 where
$\norm{\cdot}$ denotes the Frobnius norm ($\ell^2$-norm).
This implies that \textit{the linear system, i.e. the QKDE described by Eq.(\ref{eq-QKDE}), is critically (neutrally) stable and the numerical integration is sensitive to the computational errors}. Therefore,
it is necessary to find a robust and long time precise integration method for solving Eq.(\ref{eq-QKDE}). Many researchers have
studied this problem with the traditional finite difference method since 1970s. Hrastar \cite{Hrastar1970}, Cunningham \cite{Cunningham1980} and Wie \cite{Wie1985}  used the Taylor series method, Miller \cite{Miller1983} tried the rotation vector concept, Mayo \cite{Mayo1979} adopted the Runge-Kutta \cite{RK-FDS} and the state transition matrix method and Wang \cite{Wang2007} compared the Runge-Kutta scheme with symplectic difference scheme, and Funda et al. \cite{Funda1990} used the periodic normalization to unit magnitude. However, even for the time-invariant $\vec{\omega}$, the numerical scheme for QKDE
may be sensitive to the accumulative computational errors and may also encounter the stiff problem.

We solve the QKDE Eq.(\ref{eq-QKDE}) via symplectic geometric algorithms (SGA)  which overcomes the disadvantages of the previous methods.
Firstly, we consider the autonomous QKDE (A-QKDE) in which the parameters $\omega_1, \omega_2$ and $\omega_3$ are time-invariant constants. In such a scenario, the QKDE can be modeled by an autonomous Hamiltonian system and the SGA, which was firstly proposed by K. Feng \cite{Ruth1983} and R. D. Ruth \cite{Ruth1983} independently  and developed by other researchers in the past 30 years \cite{Feng1991,Kong2007,Feng2010,Sanz-Serna-1994,Hairer2006}, and can be applied directly to get a non-dissipative numerical
scheme. Secondly, we discuss the non-autonomous QKDE (NA-QKDE) where $\omega_1(t), \omega_2(t), \omega_3(t)$
depend on time explicitly and give its numerical scheme by utilizing symplectic geometric method.

The main purpose of this paper is to propose SGAs  for solving the general QKDE while preserving long time precision and the norms of quaternions interested automatically. The contents of this paper are organized logically. The preliminaries of SGA are presented in Section \ref{section2}. Section \ref{section3} deals with the SGA for the A-QKDE. In Section \ref{section4} we cope with the SGA for NA-QKDE.  The simulation results are presented in
Section \ref{section5}. Finally, Section \ref{section6} gives the summary and conclusions.

\section{Preliminaries}\label{section2}

\subsection{Hamiltonian System}

W. R. Hamiltonian introduced the canonical differential equations \cite{Arnold1989,Arnold2006}
\begin{equation*}
\fracode{p_i}{t} = -\fracpde{H}{q_i}, \quad \fracode{q_i}{t} = \fracpde{H}{p_i}, \quad i =1, 2, \cdots, N
\end{equation*}
for problems of geometrical optics, where $p_i$ are the generalized momentums,
$q_i$ are the generalized displacements and $H = H(p_1, \cdots, p_N, q_1, \cdots, q_N)$ is the Hamiltonian, viz., the total energy of the system. Let
$\vec{p} =\trsp{[p_1, \cdots, p_N]}\in \ES{R}{N}{1}$,
$\vec{q} = \trsp{[q_1, \cdots, q_N]}\in \ES{R}{N}{1}$, and
$\vec{z} =\trsp{[p_1, \cdots, p_N, q_1, \cdots, q_N]}= \trsp{[\trsp{\vec{p}}, \trsp{\vec{q}}]}\in \ES{R}{2N}{1}$,
then $H = H(\vec{p}, \vec{q}) = H(\vec{z})$ can be specified by  $\vec{z}$ in the $2N$-dimensional phase space. Since the gradient of $H$ is
 $$
H_{\vec{z}}
=\trsp{\left[ \fracpde{H}{p_1}, \cdots, \fracpde{H}{p_N}, \fracpde{H}{q_1}, \cdots, \fracpde{H}{q_N}\right]} \in \ES{R}{2N}{1}.
$$
Then the canonical equation is equivalent to
\begin{align} \label{eq-Heq}
\fracode{\vec{z}}{t} =  \inv{\mat{J}} \cdot H_{\vec{z}}(\vec{z}), \quad
\mat{J} = \begin{bmatrix} \mat{O}_N & \mat{I}_N \\ -\mat{I}_N & \mat{O}_N\end{bmatrix}
\end{align}
where $\mat{I}_N$ is the $N$-by-$N$ identical matrix, $\mat{O}_N$ is the $N$-by-$N$ zero matrix and $\mat{J}$
is the $2N$-th order standard symplectic matrix \cite{Feng1984,Feng2010}. Any system which can be described by Eq.(\ref{eq-Heq}) is called an Hamiltonian system. The canonical equation of Hamiltonian system is invariant under the symplectic transform (phase flow), the evolution of the system is the evolution of symplectic transform, both the symplectic symmetry and the total energy of the system can be preserved simultaneously and automatically \cite{Arnold1989, Kong2009, Franco2013, Hernandez2016}.

\subsection{Transition Mapping}
The symplectic geometric algorithms are motivated by these fundamental facts. Ruth \cite{Ruth1983} and Feng \cite{Feng1984}  emphasized two key points in their pioneer works:
\begin{itemize}
\item[(a)] symplectic geometric algorithm is a kind of difference scheme which preserves the symplecitc structure of Hamiltonian system;
\item[(b)]  the single-step transition mapping  is a symplectic transform (matrix) which preserves the symplectic structure of the difference equation obtained by discritizing the original continuous Eq.(\ref{eq-Heq}).
\end{itemize}
When the
initial condition
$\left.\vec{z}(t)\right|_{t_0} = \vec{z}(t_0) = \trsp{[p_1(t_0), \cdots, p_N(t_0), q_1(t_0), \cdots, q_N(t_0)]}$
is given, the symplectic difference scheme (SDS) for Eq.(\ref{eq-Heq}) can be written by
\begin{equation}
\vec{z}[k+1] = \mat{G}_\tau \vec{z}[k],\quad \vec{z}[0] = \vec{z}(t_0),  \quad k = 0, 1, 2, \cdots
\end{equation}
where $\tau$ is the time step, $\vec{z}[k] = \vec{z}(t_0+k\tau)$ is the sample value at the discrete time $t_k = t_0+k \tau$,
and
$ \mat{G}_\tau:   \ES{R}{2N}{1}\to \ES{R}{2N}{1}$, $ \vec{z}[k]  \mapsto \vec{z}[k+1] $
is the transition operator, also named as transition mapping or matrix, from $k$-th step to $(k+1)$-th step such that
\begin{equation}
\trsp{\mat{G}_\tau} \mat{J} \mat{G}_\tau = \mat{J}.
\end{equation}
Equivalently, we have
$
\mat{G}_\tau  \in \mathrm{Sp}(2N, \mathbb{R}) \subset \GL{2N}{\mathbb{R}} \subset \ES{R}{2N}{2N},
$
in which $\mathrm{Sp}(2N, \mathbb{R})$ is the symplectic transform group and $\GL{2N}{\mathbb{R}}$ is the general linear transform group \cite{Guillemin1990,Sternberg2012,Cannas2008}.
Usually, $\mat{G}_\tau$ is specified by numerical schemes adopted.

For the general nonlinear and nonseparable Hamiltonian system, a useful symplectic difference scheme, i.e., the  \textit{centered Euler implicit scheme} of the second order (CEIS-2) \cite{Feng2010}, described by
\begin{equation} \label{eq-CIES}
\begin{split}
\frac{\vec{p}[k+1] -\vec{p}[k]}{\tau}  = -  H_{\vec{q}}(\bar{\vec{z}}_k), \quad
\frac{\vec{q}[k+1] - \vec{q}[k]}{\tau} =   H_{\vec{p}}(\bar{\vec{z}}_k),
\end{split}
\end{equation}
in which the midpoint
\begin{equation}
 \bar{\vec{z}}_k = \frac{\vec{z}[k+1] +\vec{z}[k]}{2}
\end{equation}
can be used to determine the transition matrix $\mat{G}_\tau$.
Let
\begin{align}\label{eq-B}
\mat{B} = \inv{\mat{J}} H_{\vec{zz}}(\bar{\vec{z}}_k),\quad
\phi(\lambda) = \frac{1+\lambda}{1-\lambda},  %\label{eq-phi}
\end{align}
where $H_{\vec{z}\vec{z}}(\cdot) = \left(\fracpdemix{H(\cdot)}{z_i}{z_j} \right)_{2N \times 2N}$ is the Hessel matrix at the midpoint $\bar{\vec{z}}_k$,
and
$\phi(\cdot)$ is the Cayley transform. Then for small $\tau$ such that $\mat{I} - \frac{\tau}{2} \mat{B}$ is nonsingular, the $\mat{G}_\tau$ will be \cite{Feng2010}
\begin{equation}   \label{eq-Gt}
\begin{split}
\mat{G}_\tau
         = \phi\left(\frac{\tau}{2}\mat{B} \right)
         = \left[ \mat{I} + \frac{\tau}{2}\mat{B}\right]\cdot \inv{\left[ \mat{I} - \frac{\tau}{2}\mat{B}\right]}
         = \inv{\left[ \mat{I} - \frac{\tau}{2}\mat{B}\right]}\left[ \mat{I} + \frac{\tau}{2}\mat{B}\right].
\end{split}
\end{equation}
Note that although Eq.(\ref{eq-Gt}) is an approximate result  in the sense of approximate
conservation \cite{Feng1984}
for the
general nonlinear Hamiltonian system (Linear Hamiltonian system requires that the Hessel matrix $H_{\vec{zz}}(\cdot)$ is symmetric), it could be a precise solution for some special cases.

\section{Symplectic Algorithm for A-QKDE}\label{section3}
\subsection{A-QKDE and Autonomous Hamilton System}

When the matrix $\mat{A}$ in Eq.(\ref{eq-QKDE}) is time-invariant, or equivalently the parameters $p, q$ and $r$ are constants, we can model it with the autonomous Hamilton system. Obviously, for $N = 2$, let $\vec{p} = \trsp{[p_1, p_2]} = \trsp{[e_0, e_1]}$,
 $\vec{q}=\trsp{[q_1, q_2]} = \trsp{[e_2, e_3]}$, $\quat{q} = \trsp{[\trsp{\vec{p}}, \trsp{\vec{q}}]}\equiv \vec{z}$, we have
\begin{equation} \label{eq-Q2HS}
\fracode{\vec{z}}{t} = \inv{\mat{J}}H_{\vec{z}}(\vec{z}) = \frac{1}{2}\mat{A}(\vec{\omega}) \vec{z}
\end{equation}
with the help of Eq.(\ref{eq-QKDE}) and Eq.(\ref{eq-A}). We remark here that the symbols $\vec{z}$
and $\quat{q}$ are equivalent since the vector $\vec{z}$ consists of $e_0, e_1, e_2, e_3$. Thus $\quat{q}$ and $\vec{z}$ can be used alternatively if necessary.   From Eq.(\ref{eq-Q2HS}) we can find that
$ \inv{\mat{J}}H_{\vec{z}}(\vec{z}) =\frac{1}{2}\mat{A} \vec{z}$.
Therefore
\begin{equation} \label{eq-Bq}
\begin{split}
\mat{A} = 2\inv{\mat{J}}H_{\vec{zz}}(\vec{z}),\quad
H_{\vec{zz}}(\vec{z}) = \frac{1}{2}\mat{J}\mat{A},\quad
\mat{B} = \inv{\mat{J}}H_{\vec{zz}}(\vec{z}) = \frac{1}{2}\mat{A}
 \end{split}
\end{equation}
according to
Eq.(\ref{eq-B}). Note that  $\mat{J}\mat{A}\neq \mat{A}\mat{J}$, so  the Hessel matrix
$H_{\vec{zz}}(\vec{z})$ is not symmetric, which means that the Hamiltonian system here is \textit{nonlinear} by definition \cite{Feng2010}.

\subsection{Symplectic Transition Mapping}

Obviously, the A-QKDE is an autonomous Hamiltonian system and the
symplectic algorithm can be adopted to find its numerical solution. In the following,
 we will prove a lemma and an interesting Euler's formula for constructing the symplectic difference scheme for A-QKDE.
\begin{lem} \label{lem-1}
Suppose matrix $\mat{M}\in \ES{R}{2n}{2n}$ is skew-symmetric, i.e., $\trsp{\mat{M}} = -\mat{M}$ and there exists
a positive constant $\gamma$ such that $\mat{M}^2 = -\gamma^2 \mat{I}$. Let $\phi(\lambda) = \frac{1+\lambda}{1-\lambda}$ be the
Caylay transformation and $\hat{\mat{M}} = {\gamma}^{-1}\mat{M}$, then for any $x\in \mathbb{R}$ the Euler formula
\begin{equation}
\begin{split}
\phi(x\mat{M})
= \frac{1}{1+\alpha}[(1-\alpha)\mat{I} +2x\mat{M}]
=\cos \theta(x, \gamma) \mat{I} + \sin \theta(x, \gamma) \hat{\mat{M}}, \quad \hat{\mat{M}}^2 = -\mat{I}
\end{split}
\end{equation}
holds, in which $\theta = \theta(x, \gamma) = 2\cdot \arctan (x\gamma)$ and $\alpha = x^2\gamma^2$. Furthermore, $\phi(x\mat{M})$ is an orthogonal matrix.
\end{lem}

\textsc{Proof}: It is trivial that $\trsp{\hat{\mat{M}}} = -\hat{\mat{M}}$ and $\hat{\mat{M}}^2 = -\mat{I}$. For any $x\in \mathbb{R}$, we find that
$(\mat{I}-x\mat{M})(\mat{I}+x\mat{M})= (1+x^2\gamma^2)\mat{I}$. In consequence
$\inv{(\mat{I}-x\mat{M})} = \frac{1}{1+x^2\gamma^2}(\mat{I}+x\mat{M})$.
Hence the Caylay transformation $\phi(x\mat{M})$ can be simplified as following
\begin{equation} \label{eq-phi-xM}
\begin{split}
\phi(x\mat{M}) &= \inv{(\mat{I}-x\mat{M})} (\mat{I}+x\mat{M})
= \frac{1}{1+\alpha}[(1-\alpha)\mat{I} + 2x\mat{M}]
 \end{split}
\end{equation}
where $\alpha = x^2\gamma^2$. Put $\tan \frac{\theta}{2} = x\gamma, \alpha = x^2\gamma^2$,
then by the trigonometric identity and $\mat{M} = \gamma\hat{\mat{M}}$ we immediately have
\begin{equation}
\begin{split}
\phi(x\mat{M})
 = \frac{1}{1+x^2\gamma^2}[(1-x^2\gamma^2)\mat{I} + 2x\gamma \hat{\mat{M}}]
 &= \cos \theta(x,\gamma) \mat{I} + \sin \theta(x,\gamma) \hat{\mat{M}}.
\end{split}
\end{equation}
Moreover,
\begin{align*}
\trsp{[\phi(x\mat{M})]}\cdot \phi(x\mat{M})
&= \trsp{[\cos \theta(x,\gamma) \mat{I} + \sin \theta(x,\gamma) \hat{\mat{M}}]}[\cos \theta(x,\gamma) \mat{I} + \sin \theta(x,\gamma) \hat{\mat{M}}] \\
& = \cos^2 \theta(x,\gamma) \mat{I} - \sin^2 \theta(x,\gamma) \hat{\mat{M}}^2
= \mat{I}.
\end{align*}
Similarly, we can obtain $\phi(x\mat{M})\cdot\trsp{[\phi(x\mat{M})]} = \mat{I}$. Hence $\phi(x\mat{M})$ is orthogonal by definition.
%%%%%%%%%%%%%%
%% Theorem 1
%%%%%%%%%%%%%%
\begin{thm} (\textbf{Euler's formula}) For any time step $\tau$ and time-invariant vector $\vec{\omega}$, the transition mapping $\mat{G}_\tau^{\quat{q}}$ for the A-QKDE can be represented by
\begin{equation}  \label{eq-GtqExp}
\begin{split}
\mat{G}_\tau^{\quat{q}}
= \frac{1}{1+\alpha}  \left[ \left( 1 - \alpha\right)\mat{I}
    +\frac{\tau}{2}\mat{A}\right]
= \cos\theta(\vec{\omega}, \tau)  \cdot \mat{I} + \sin \theta(\vec{\omega},\tau) \cdot \hat{\mat{A}},
\end{split}
\end{equation}
in which $\alpha = \tau^2\norm{\vec{\omega}}^2/16$, $\theta = 2\arctan(\tau\norm{\vec{\omega}}/4)$, $\hat{\mat{A}}=\mat{A}/\norm{\vec{\omega}}$ and $\hat{\mat{A}}^2 = -\mat{I}$.
\end{thm}

\noindent \textsc{Proof}: With the help of Eq.(\ref{eq-Gt}) and Eg.(\ref{eq-Bq}), the transition mapping will be
$
\mat{G}_\tau^{\quat{q}} = \phi\left(\frac{\tau}{2}\mat{B}_{\vec{z}}\right) = \phi\left(\frac{\tau}{4}\mat{A}\right)
$.
Let $x=\tau/4, \mat{M} = \mat{A}, \gamma = \norm{\vec{\omega}}$,
then $\alpha = x^2 \gamma^2 =\norm{\vec{\omega}}^2\tau^2/16$. Thus  the theorem follows from
Lemma \ref{lem-1} directly.

%%%%%%%%%%%%%%%%%%%%%%%%%%%%%%%%
%%  Theorem ２
%%%%%%%%%%%%%%%%%%%%%%%%%%%%%%%%

\begin{thm}
For any $\vec{\omega}\in \ES{R}{3}{1}$ and $\tau\in \mathbb{R}$, the transition mapping $\mat{G}_\tau^{\quat{q}}$
is an orthogonal transformation and an invertible symplectic transformation with first-order precision, i.e.,
\begin{align}\label{eq-Gq-1st-sp}
\trsp{[\mat{G}_\tau^{\quat{q}}]} \mat{J} \mat{G}_\tau^{\quat{q}}  = \mat{J} + \mathcal{O}(\tau), \quad
\inv{[\mat{G}_\tau^{\quat{q}}]} = \mat{G}_{-\tau}^{\quat{q}} = \trsp{[\mat{G}_\tau^{\quat{q}}]}.
\end{align}
\end{thm}

\noindent \textsc{Proof}: For a constant vector $\vec{\omega}$, we can find the function $ \theta = 2\arctan (\tau \norm{\vec{\omega}}/4)$ is an odd function of time step $\tau$. By utilizing $ \hat{\mat{A}}^2 = -\mat{I}$ and $\trsp{\hat{\mat{A}}} = -\hat{\mat{A}}$, we immediately obtain
\begin{align*}
\trsp{[\mat{G}_\tau^{\quat{q}}]}
&= \trsp{[\cos \theta(\vec{\omega},\tau)\mat{I} +\sin \theta(\vec{\omega},\tau)\hat{\mat{A}}]}\\
&= \cos \theta(\vec{\omega},\tau)\mat{I} -\sin \theta(\vec{\omega},\tau)\hat{\mat{A}}
= \cos (-\theta(\vec{\omega},\tau))\mat{I} +\sin (-\theta(\vec{\omega},\tau))\hat{\mat{A}}\\
&= \cos \theta(\vec{\omega},-\tau))\mat{I} +\sin \theta(\vec{\omega},-\tau)\hat{\mat{A}}
= \mat{G}_{-\tau}^{\quat{q}},
\end{align*}
which implies that the transition mapping is invertible. Moreover, $\mat{G}_{\tau}^{\quat{q}}$ is orthogonal by
Lemma \ref{lem-1}. In consequence, $\trsp{[\mat{G}_\tau^{\quat{q}}]} = \mat{G}_{-\tau}^{\quat{q}} = \inv{\left(\mat{G}_{\tau}^{\quat{q}}\right)}$.
Furthermore, simple algebraic operation implies that
$\mat{J}\hat{\mat{A}} -\hat{\mat{A}}\mat{J}= 2 (\mat{J} + \hat{q} \hat{\mat{A}}) \hat{\mat{A}}$
because $\hat{\mat{A}}\mat{J}\hat{\mat{A}} = \mat{J} + 2\hat{q} \hat{\mat{A}}$ and
$ \inv{\hat{\mat{A}}} = -\hat{\mat{A}}$,
where $\hat{q} =\omega_2/\norm{\vec{\omega}}$. Therefore,
\begin{align*}
\trsp{[\mat{G}_\tau^{\quat{q}}]} \mat{J} [\mat{G}_\tau^{\quat{q}}]
= \mat{J} + ( \mat{J} +\hat{q}\hat{\mat{A}} )\cdot(-2\sin^2\theta \mat{I} + 2\sin\theta \cos \theta \hat{\mat{A}}).\
\end{align*}
Since both $\hat{q}$ and $\hat{\mat{A}}$ are independent with $\tau$, $
\sin\theta = \frac{2\tan\frac{\theta}{2}}{1+\tan^2\frac{\theta}{2}}
= \frac{2\cdot \frac{1}{4}\norm{\vec{\omega}}\tau}{1+\frac{1}{16}\norm{\vec{\omega}}^2\tau^2}$, $
\cos\theta = \frac{1-\tan^2\frac{\theta}{2}}{1+\tan^2\frac{\theta}{2}}
= \frac{1+\frac{1}{16}\norm{\vec{\omega}}^2\tau^2}{1+\frac{1}{16}\norm{\vec{\omega}}^2\tau^2}$,
we can deduce that
$ \trsp{[\mat{G}_\tau^{\quat{q}}]} \mat{J} [\mat{G}_\tau^{\quat{q}}] = \mat{J} + \mathcal{O}(\tau)$.
Hence  $\mat{G}_\tau^{\quat{q}}$ is a symplectic matrix with first-order precision by definition\cite{Feng2010}.

We remark that the nonlinearity of the QKDE, or equivalently the non-symmetric property of the Hessel matrix $\nabla^2 H(\quat{q})=H_{\vec{zz}}(\vec{z})$,
brings the non-commutativity of $\mat{J}$ and $\hat{\mat{A}}$, i.e., $\mat{J}\hat{\mat{A}} \neq \hat{\mat{A}}\mat{J}$, \textit{which
lowers the precision of the Euler implicit midpoint formula  from the second order to the first order}.

\subsection{Comparison with Analytic Solution}

Fortunately, the analytic solution (AS) for A-QKDE can be found without difficulty.
In fact for any $t\in [t_0,t_f]$ where $t_0$ and $t_f$ denote the initial and final time respectively, we have
\begin{equation}
\quat{q}(t) = \exp{\left(\frac{1}{2}\mat{A}\cdot(t - t_0)\right)} \cdot \quat{q}(t_0), \quad t\in[t_0, t_f].
\end{equation}
Let $x = (t-t_0)/2$ and $\hat{\mat{A}}
=\mat{A}/\norm{\vec{\omega}}$, then $\mat{A}^2 = -\norm{\vec{\omega}}^2 \mat{I}$ show that
\begin{align*}
\exp{(\mat{A}x)}
&= \sum^\infty_{k=0} \frac{(\mat{A}x)^{2k}}{(2k)!} + \sum^\infty_{k=0} \frac{(\mat{A}x)^{2k+1}}{(2k+1)!}
= \cos(\norm{\vec{\omega}}x)\mat{I} + \sin(\norm{\vec{\omega}}x)\hat{\mat{A}} \\
&= \cos\left(\frac{1}{2}\norm{\vec{\omega}}(t-t_0) \right)\mat{I}
  + \sin \left(\frac{1}{2}\norm{\vec{\omega}}(t-t_0) \right) \hat{\mat{A}}.
\end{align*}
Thus for $t-t_0 = \tau$, the AS to the transition mapping for the A-QKDE is
\begin{equation}
\mat{\mat{G}}^{\quat{q}}_{\tau,\mathrm{AS}}
 =  \mat{I} \cos \frac{\norm{\vec{\omega}}\tau}{2}
  + \hat{\mat{A}} \sin \frac{\norm{\vec{\omega}}\tau}{2}  , \quad \forall \tau \in \mathbb{R}.
\end{equation}
At the same time, for sufficiently small $\tau$ in Eq.(\ref{eq-GtqExp}) we have
\begin{equation}
\begin{split}
\mat{\mat{G}}^{\quat{q}}_{\tau}
=\mat{I} \cos(2 \arctan \frac{\norm{\vec{\omega}}\tau}{4} )
  + \hat{\mat{A}} \sin (2 \arctan \frac{\norm{\vec{\omega}}\tau}{4} )
\sim \mat{I} \cos \frac{\norm{\vec{\omega}}\tau}{2}
  +  \hat{\mat{A}} \sin \frac{\norm{\vec{\omega}}\tau}{2}
\end{split}
\end{equation}
for sufficiently small $x=\norm{\vec{\omega}\tau}$ since
\begin{align*}
\cos\frac{x}{2} - \cos\left(2\arctan\frac{x}{4}\right)
&= -\frac{x^4}{192} + \frac{43x^6}{92160}- \frac{157x^8}{5160960}
+\frac{14173x^{10}}{7431782400} + \mathcal{O}(x^{12}),\\
\sin\frac{x}{2} - \sin\left(2\arctan\frac{x}{4}\right)
&= \frac{x^3}{96}-\frac{13x^5}{7680} + \frac{311x^7}{2580480} - \frac{2833x^9}{371589120} +  \mathcal{O}(x^{11}).
\end{align*}
Let $h(x) =\max\set{\abs{\cos\frac{x}{2} - \cos\left(2\arctan\frac{x}{4}\right)}, \abs{\sin\frac{x}{2} - \sin\left(2\arctan\frac{x}{4}\right)}}$, then when we have $ h(x) < 1.25\times 10^{-4}$ for $x  \le 0.2 $ and $h(x) < 1.57\times 10^{-8}$ for $x \le 0.01$.
Therefore, the $\mat{\mat{G}}^{\quat{q}}_{\tau,AS}$  can be approximated by $\mat{\mat{G}}^{\quat{q}}_{\tau}$ with an acceptable precision when time step $\tau \le 1/(5\norm{\vec{\omega}})$.

\subsection{Symplectic Geometric Algorithm}
The SAG for A-QKDE is given in Algorithm \ref{Sp-A-QKDE} by the explicit expression $G^{\quat{q}}_\tau$  obtained.
\begin{algorithm}[h] \label{Sp-A-QKDE}
\caption{\textbf{Symplectic Geometric Scheme for A-QKDE with CEIS-2}}
\begin{algorithmic}[1]
  \REQUIRE The time-invariant vector $\vec{\omega}\in \ES{R}{3}{1}$, time step $\tau$ such
  that $\tau \le \frac{1}{5\norm{\vec{\omega}}}$ and the
           initial quaternion $\quat{q}_0 = \trsp{[e_0(t_0), e_1(t_0), e_2(t_0), e_3(t_0)]}$ at initial time $t_0$.
  \ENSURE  Numerical solution to the A-QKDE
     $\fracode{\quat{q}}{t} = \frac{1}{2}\mat{A}(\vec{\omega})\quat{q}$ for $ t \geq t_0$
           with first-order symplectic difference scheme.
  \STATE Set matrix $\mat{A}$ with vector $\vec{\omega}=\trsp{[\omega_1, \omega_2, \omega_3]}$ according to Eq.(\ref{eq-A}).
  \STATE Set parameter $\alpha$ with $\vec{\omega}$ and $\tau$, viz.
          $ \alpha = \frac{1}{16}\tau^2\norm{\vec{\omega}}^2 =\frac{1}{16}\tau^2(\omega_1^2 + \omega_2^2 + \omega_3^2) $.
  \STATE Compute the transition mapping:
          $\mat{G}^{\quat{q}}_\tau = \frac{1}{1+\alpha}\left[ (1-\alpha)\mat{I} + \frac{\tau}{2}\mat{A}\right]$.
  \STATE Set the initial condition $\quat{q}[0] =\quat{q}(t_0)= \quat{q}_0$
  \STATE Iterate: $\quat{q}[k+1] = \mat{G}^{\quat{q}}_\tau \quat{q}[k], \quad k = 0, 1, 2, \cdots $
\end{algorithmic}
\end{algorithm}

\section{Symplectic Geometric Method for NA-QKDE}\label{section4}

\subsection{Time-dependent Parameters and NA-QKDE}

For the time-varying vector $\vec{\omega} = \vec{\omega}(t)$, Eq.(\ref{eq-QKDE}) implies
$
\dot{\vec{z}}=\inv{\mat{J}}H_{\vec{z}}(\vec{z}) = \frac{1}{2}\mat{A}(\vec{\omega}(t)) \vec{z}
$,
where $H = H(p_1, p_2, q_1,q_2,t)= H(e_0, e_1, e_2, e_3, t)$. Hence the NA-QKDE is a non-autonomous Hamiltonian system essentially and the corresponding symplectic geometric algorithm  can be obtained via the concept of extended phase space \cite{Feng2010} (See Chapter 5, Section 5.8).
Let $p_3 = h, q_3=t$, where $h$ is the negative
of the total energy. Let
$\vec{z}=\trsp{[p_1, p_2, h, q_1, q_2, t]}
= \trsp{[\trsp{\vec{p}}, h, \trsp{\vec{q}}, t]}\in \ES{R}{(2N+2)}{1}$,
$K(\tilde{\vec{z}})= h + H(p_1, p_2, q_1, q_2,t)
= h + H(\vec{p},\vec{q},t) = h + H(\quat{q},t)$,
then we have \textit{the time-centered Euler
implicit scheme of the fourth order} (T-CEIS-4) \cite{Feng2010},
\begin{equation} \label{eq-CIES2}
\frac{\vec{z}[k+1] -\vec{z}[k]}{\tau}  = \mat{J}^{-1}H_{\vec{z}}(\bar{\vec{z}}_k) - \frac{\tau^2}{24}\mat{J}^{-1}\nabla_{\vec{z}}\left\{\trsp{[H_{\vec{z}}(\bar{\vec{z}}_k)]} \mat{J}^{-1}H_{\vec{zz}}(\bar{\vec{z}}_k)\mat{J}^{-1}H_{\vec{z}}(\bar{\vec{z}}_k)\right\}.
\end{equation}
Note that in our problem we have no interest in the concrete results for $h$. Eq.(\ref{eq-CIES2}) shows that for the time-varying problems,  $H_*(\bar{\vec{z}}_k)$ is replaced by $H_*(\bar{\vec{z}}_k, \bar{t}_k)$ where $*$ may be $\vec{p}, \vec{q}$ or $t$. In order to guarantee the duality of the phase space, $\vec{z}$ is rewritten as $\vec{z}= \trsp{[\trsp{\vec{p}}, \trsp{\vec{q}}]}$.
The symplectic scheme for the
NA-QKDE as Eq.(\ref{eq-CIES2}) can be simplified as
\begin{equation} \label{eq-CIES-qkde}
\frac{\vec{z}[k+1] -\vec{z}[k]}{\tau}  = \frac{1}{2}\mat{A}_k\bar{\vec{z}}_k - \frac{\tau^2\Omega_k\norm{\vec{\omega}_k}^2}{96}\mat{J}\bar{\vec{z}}_k,
\end{equation}
where $\bar{t}_k = t[k] +\tau/2$,  $\vec{\omega}_k =\vec{\omega}(\bar{t}_k) $, $\mat{A}_k = \mat{A}(\vec{\omega}_k)$ and
\begin{equation}\label{eq:T-CIES-4-Omega}
\Omega_k = \omega_2(\bar{t}_k).
\end{equation}
By taking the similar procedure as we do
in finding the transition mapping for symplectic difference scheme of the A-QKDE, we can obtain
\begin{equation}
\mat{G}_{\tau}^{\quat{q}}(k+1|k) =
\inv{\left[ \mat{I} - \frac{\tau}{2}\mat{B}_k\right]}\left[ \mat{I} + \frac{\tau}{2}\mat{B}_k\right]
\end{equation}
where $\mat{B}_k$ is a skew-symmetric matrix such that
\begin{equation}\label{eq:na-qkde-b}
\mat{B}_k = \frac{1}{2}\mat{A}_k - \frac{\tau^2\Omega_k\norm{\vec{\omega}_k}^2}{96}\mat{J}.
\end{equation}
According to Lemma.\ref{lem-1}, there exists a series of $\gamma_k$ which satisfy the following identity
\begin{equation}
\mat{B}_k = - \gamma_k^2\mat{I}.
\end{equation}
\begin{thm}
For the NA-QKDE~$\fracode{\quat{q}}{t} = \frac{1}{2}\mat{A}[\vec{\omega}(t)]\quat{q}$, let $\beta_k = -\frac{\Omega_k^2}{96}\norm{\vec{\omega}_k}^2$,
 $\gamma_k^2 = \frac{1}{4}\norm{\vec{\omega}_k}^2-\beta_k\Omega_k  + \beta_k^2$,
$\alpha_k = \frac{\tau^2}{4}\gamma_k^2 = \frac{1}{16}\tau^2 \norm{\vec{\omega}_k}^2(1+\frac{\tau^2\Omega_k^2}{24}++\frac{\tau^4\Omega_k^2\norm{\vec{\omega}_k}^2}{2304})$, $\hat{\mat{B}}_k = \frac{1}{\gamma_k}\mat{B}_k$,
 then the transition mapping
$ \mat{G}^{\quat{q}}_\tau(k+1|k): \vec{z}[k] \mapsto \vec{z}[k+1], k =0, 1, 2, \cdots $
can be given by
\begin{equation}
\begin{split}
\mat{G}^{\quat{q}}_\tau(k+1|k)
= \frac{1}{1+\alpha_k}\left[ (1-\alpha_k)\mat{I} + \tau\mat{B}_k\right]
= \cos \theta\left(\vec{\omega}_k,\tau\right)  \mat{I}
  +\sin \theta\left(\vec{\omega}_k,\tau\right)  \hat{\mat{B}}(\vec{\omega}_k),
\end{split}
\end{equation}
where $\theta = 2\arctan\frac{\tau\norm{\vec{\omega}_k}}{4}\sqrt{1+\frac{\tau^2}{24}\Omega_k^2+\frac{\tau^4}{2304}\Omega_k^2\norm{\vec{\omega}_k}^2}$ such that
\begin{equation} \label{eq-Gk-1st-sp}
\trsp{[\mat{G}^{\quat{q}}_\tau(k+1|k)]}\cdot \mat{J}\cdot [\mat{G}^{\quat{q}}_\tau(k+1|k)] = \mat{J} +\mathcal{O}(\tau^2).
\end{equation}
\end{thm}
Due to the nonlinearity of the QKDE, \textit{the precision of T-CEIS-4 is reduced from the fourth order to the second order}.
In addition, since $\vec{\omega}(t)$ is time-dependent and $\vec{\omega}(t[k]-\tau/2) \neq \vec{\omega}(t[k]+\tau/2)$
 for positive $\tau$ in general, we can deduce that
 $\mat{G}^{\quat{q}}_{-\tau}(k+1|k) \neq \inv{[\mat{G}^{\quat{q}}_\tau(k+1|k)]}$
although $\mat{G}^{\quat{q}}_\tau(k+1|k)$ is also a symplectic and orthogonal transformation.

\subsection{Symplectic Geometric Scheme for NA-QKDE}
The SAG for NA-QKDE is presented in Algorithm \ref{Sp-NA-QKDE}.
\begin{algorithm}[h] \label{Sp-NA-QKDE}
\caption{\textbf{Symplectic Geometric Scheme for NA-QKDE with T-CEIS-4}}
\begin{algorithmic}[1]
  \REQUIRE The time-varying vector $\vec{\omega}(t)\in \ES{R}{3}{1}$band the initial quaternion
 $\quat{q}_0 = \trsp{[e_0(t_0), e_1(t_0), e_2(t_0), e_3(t_0)]}$ at initial time $t_0$.
  \ENSURE  Numerical solution to the NA-QKDE
     $\fracode{\quat{q}}{t} = \frac{1}{2}\mat{A}(\vec{\omega}(t))\quat{q}$ for $ t \geq t_0$
           with second-order symplectic difference scheme.
\STATE Set the initial condition $\quat{q}[0] =\quat{q}(t_0)= \quat{q}_0$.
  \STATE set $\bar{t}_k = t_0 + (k+1/2)\tau$.
  \STATE Set matrix $\mat{A}_k$ with vector $\vec{\omega}_k=\trsp{[\omega_1(\bar{t}_k), \omega_2(\bar{t}_k), \omega_3(\bar{t}_k)]}$ according to Eq.(\ref{eq-A}) and $\Omega_k = \omega_2(\bar{t}_k)$ by Eq.(\ref{eq:T-CIES-4-Omega}).
  \STATE Calculate the norm of $\vec{\omega}_k$:~
 $\norm{\vec{\omega}_k}^2=\omega_1(\bar{t}_k)^2 + \omega_2(\bar{t}_k)^2 + \omega_3(\bar{t}_k)^2$.
  \STATE Set parameter $\beta_k$ and $\alpha_k$ with $\vec{\omega}$ and $\tau$, viz.\\
$ \beta_k = - \frac{\tau^2}{96}\Omega_k \norm{\vec{\omega}_k}^2$,
$ \alpha_k = \frac{1}{4}\tau^2(\frac{1}{4}\norm{\vec{\omega}_k}^2- \Omega_k\beta_k  + \beta_k^2) $.
  \STATE Set matrix $\mat{B}_k$ with $\mat{\omega}_k$ and $\tau$ according to Eq.(\ref{eq:na-qkde-b}).
  \STATE Compute the transition mapping:
          $\mat{G}^{\quat{q}}_\tau[k] = \frac{1}{1+\alpha_k}\left[ (1-\alpha_k)\mat{I} + \tau\mat{B}_k\right]$.
  \STATE Iterate: $\quat{q}[k+1] = \mat{G}^{\quat{q}}_\tau[k] \quat{q}[k], \quad k = 0, 1, 2, \cdots $
\end{algorithmic}
\end{algorithm}

We remark here that $\vec{\omega}_k = \vec{\omega}(\bar{t}_k)
=\vec{\omega}(t[k]+\tau/2)$ relates the fractional interval sampling, which will increase the complexity of the
hardware implementation. However, if the sampling rate $f_s = 1/\tau$ is large enough, the  $\vec{\omega}(t)$ will vary slowly in each short time interval $[t[k], t[k+1]]$ and the linear interpolation can be considered here. This is to say that for $t\in [t[k],t[k+1]]$
\begin{equation}
\vec{\omega}(t) \approx \vec{\omega}(t[k]) + \frac{\vec{\omega}(t[k+1]) - \vec{\omega}(t[k])}{\tau} t.
\end{equation}
Hence
\begin{equation}
\vec{\omega}(\bar{t}_k) \approx \vec{\omega}(t[k]) + \frac{\vec{\omega}(t[k+1]) - \vec{\omega}(t[k])}{\tau}\cdot
\bar{t}_k
\end{equation}
with acceptable precision. In this way the fractional interval sampling can be  avoided and the complexity of the hardware implementation can be reduced.

\section{Numerical Simulation}\label{section5}

\subsection{Key Issues for Verification and Validation}

Although the principles and algorithms have been given in the previous sections, it is still necessary to
verify them with concrete examples by numerical simulation. However, each quaternion $\quat{q}$ has four
components, i.e., $e_0(t), e_1(t), e_2(t)$ and $e_3(t)$, which implies that it is not convenient for visualization unless
we project the 4-dimensional vector representation of  quaternions into a subspace. Since all of the transition mappings $\mat{G}^\quat{q}_\tau$ are orthogonal, then the norm of the quaternions should be preserved as
$\norm{\quat{q}(t)} =1$.
Thus the norm can be used as a necessary condition so as to validate the correctness of the algorithms proposed. On the other hand, we can construct some
special cases in which the analytic solutions can be obtained easily and compared with the numerical solutions.

For A-QKDE, the $\vec{\omega}$ is time-invariant and $\omega=\norm{\vec{\omega}}$ is a constant. Since the eigen-values of matrix $\mat{A}(\vec{\omega})$ are $\pm \mj \omega $, then the general solution of the A-QKDE must
be
\begin{equation}\label{eq-ei-cosine}
e_i(t) = c_i \cos(\omega t + \varphi_i), \quad t\in \mathbb{R},
\end{equation}
in which the amplitudes $c_i$ and phases $\varphi_i$ can be determined by the initial condition.

For NA-QKDE, if we choose the functions $\omega_2(t)$ and $\omega_3(t)$ such that for sufficient large $t$, $\omega_2(t)\to 0$ and $\omega_3(t)\to 0$, then Eq.(\ref{eq-QKDE}) shows that there exists some positive $t_*$ and for $t>t_*$ such that
\begin{align}
\fracode{}{t}\begin{bmatrix} e_0\\e_1\end{bmatrix}
= \begin{bmatrix} 0 &  -\frac{1}{2}\omega_1 \\ \frac{1}{2}\omega_1 & 0 \end{bmatrix}\begin{bmatrix} e_0\\e_1\end{bmatrix}, \quad \fracode{}{t}\begin{bmatrix} e_2\\e_3\end{bmatrix}
= \begin{bmatrix} 0 & \frac{1}{2}\omega_1 \\ -\frac{1}{2}\omega_1 & 0 \end{bmatrix}\begin{bmatrix} e_2\\e_3\end{bmatrix}, \quad t\geq t_*.
\label{eq-e0e1}
\end{align}
since $\quat{q}=\trsp{[e_0, e_1, e_2, e_3]}$. Hence $ e^2_0(t) + e^2_1(t) = e^2_0(0) + e^2_1(0),
e^2_2(t) + e^2_3(t) = e^2_2(0) + e^2_3(0)
$ for sufficiently large $t$.
If $\omega_1(t)\to a $ asymptotically (where $a$ is a constant), then
we have
\begin{equation} \label{eq-38}
\secode{e_i(t)}{t} + \delta^2_i  e_i(t) = 0, \quad t\geq t_*, \quad i = 0, 1, 2, 3
\end{equation}
where $\delta_i = a/2$. In other words, asymptotically,
each $e_i(t)$ can be described by Eq.(\ref{eq-ei-cosine}).

Without loss of generality, we can set the initial condition as $\quat{q}(0)=\trsp{[1,0,0,0]}$ for the verification  thus can just consider $e_0(t)$ and $e_1(t)$. Let
$ x=\tau/2$, $\bar{t}_k = t[k] + \tau/2$, $ \gamma = \omega_1(\bar{t}_k)/2$, $\mat{M} = -\gamma \mat{J}_2 $, $\hat{\mat{M}} = -\mat{J}_2$ and $\tan \frac{\theta}{2} = x\gamma = \tau \omega_1(\bar{t}_k)/4$,
then by Lemma \ref{lem-1} yields
\begin{equation}
\mat{G}^{e_0e_1}_\tau(k+1|k) = \phi(x\mat{M}) = \cos \theta \mat{I}_2 - \sin \theta \mat{J}_2
\end{equation}
where $\theta = 2\arctan \frac{\omega_1(\bar{t}_k)\tau}{4}$.
Obviously, $\mat{G}^{e_0e_1}_\tau(k+1|k)$ is a symplectic matrix since
\begin{equation}
\begin{split}
\trsp{[\mat{G}^{e_0e_1}_\tau(k+1|k)]}\cdot \mat{J}_2 \cdot [\mat{G}^{e_0e_1}_\tau(k+1|k)]
%= (\cos \theta \mat{I}_2 + \sin \theta \mat{J}_2)\mat{J}_2 (\cos \theta \mat{I}_2 - \sin \theta \mat{J}_2)
=\mat{J}_2.
\end{split}
\end{equation}

\subsection{Numerical Examples}

In order to evaluate the accuracy and stability of our proposed algorithms, we
carried out some numerical simulations of
time-variant and time-varying systems.
All experiments in this paper were implemented
in MATLAB and ran on a desktop PC equipped with Intel\textsuperscript{\textregistered} Core\textsuperscript{\texttt{TM}} i7-3770 CPU @ 3.4 GHz and 4 GB RAM.

\begin{figure}[t]
  \subfigure[$\vec{\omega}=\trsp{[2,10,3]}$]{
        \resizebox{7cm}{3cm}{\includegraphics{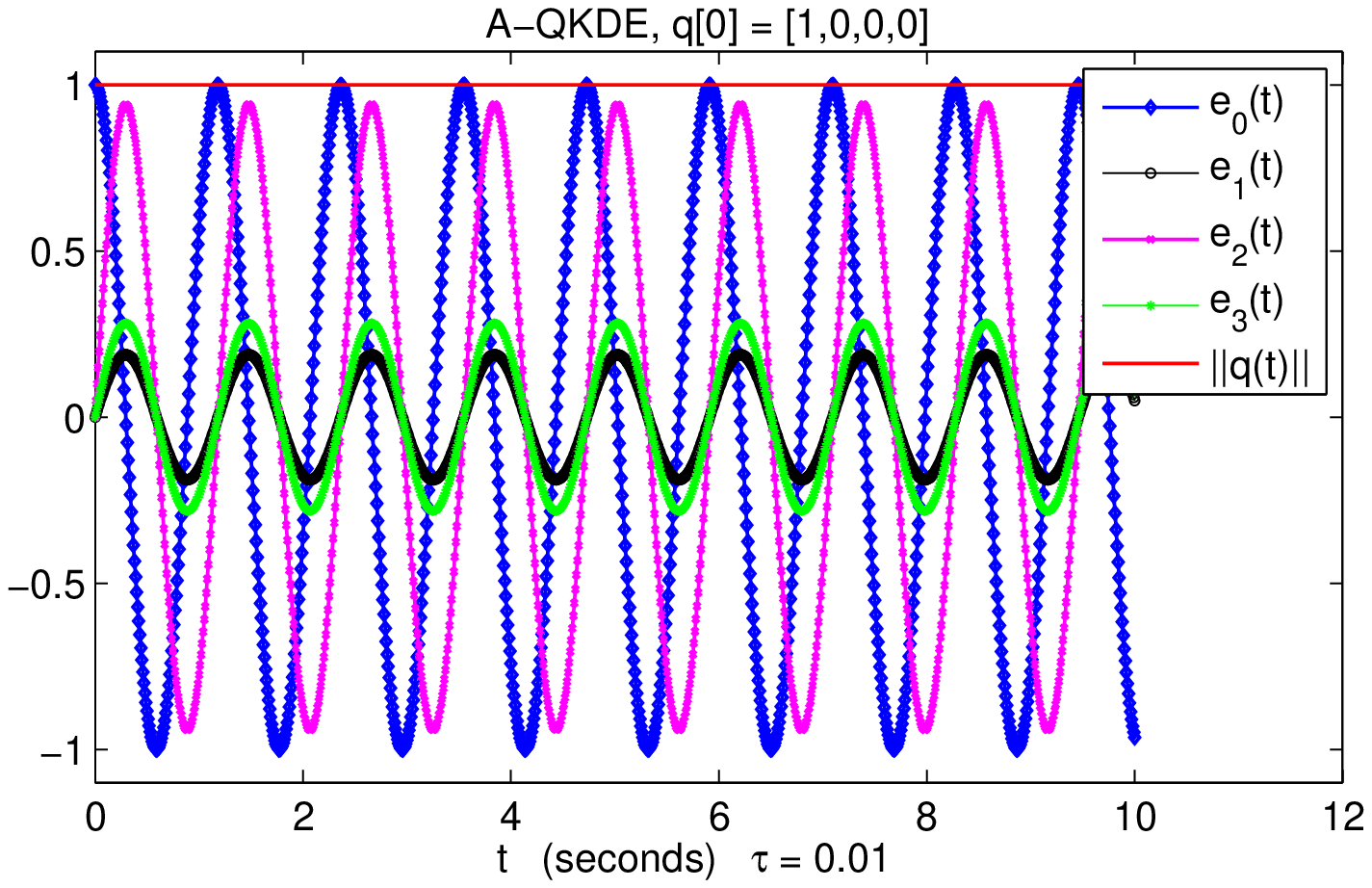}}
        \label{fig-1-Case-1}
   }
  \subfigure[$\vec{\omega}=\trsp{[2(1+\sin t\me^{-\frac{t}{4}}),0,0]}$]{
        \resizebox{7cm}{3cm}{\includegraphics{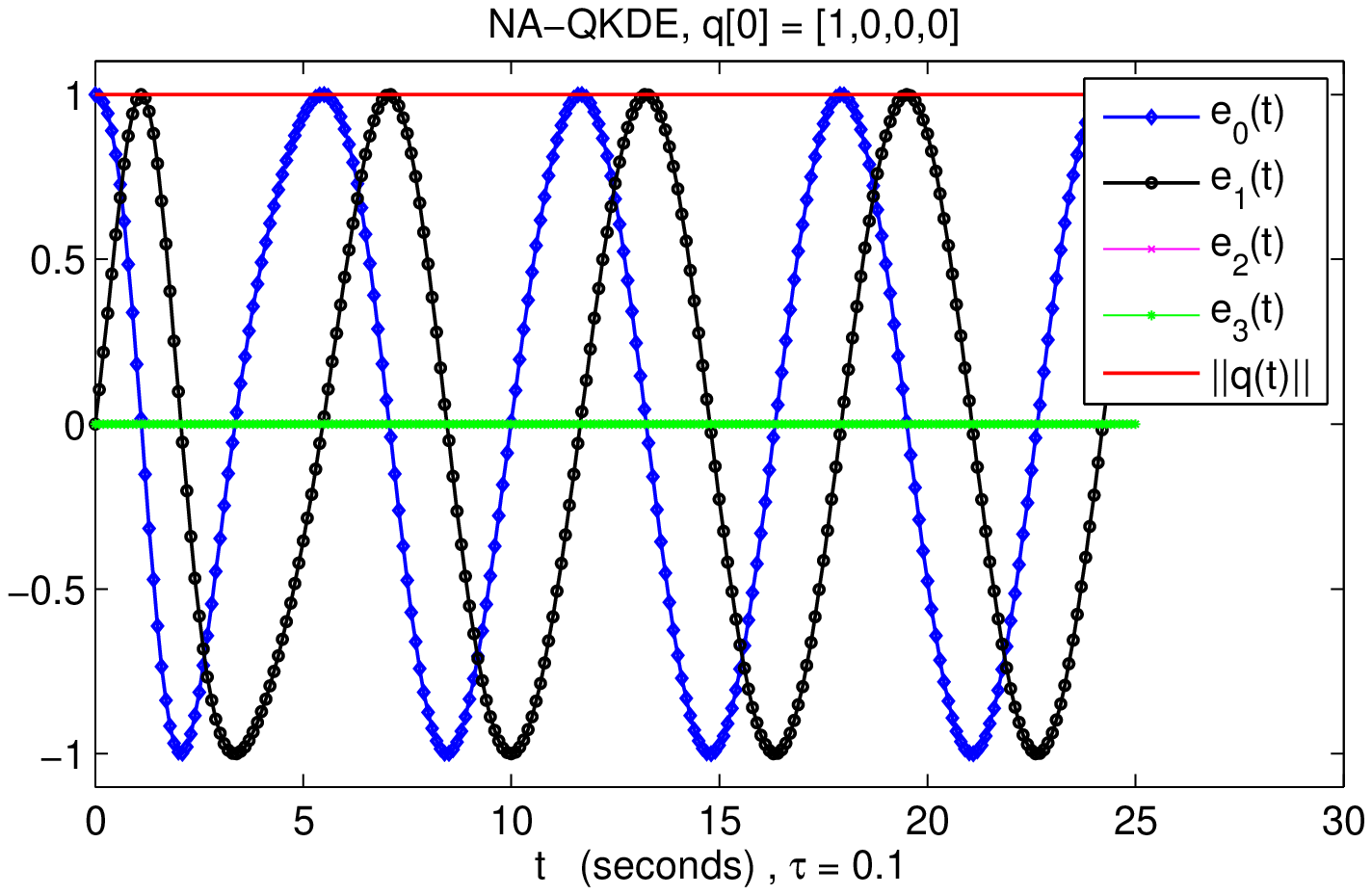}}
        \label{fig-1-Case-2}
  }
  \subfigure[$\vec{\omega}=\trsp{[2(1+\sin t\me^{\frac{-t}{4}}), (-3+t^2)\me^{\frac{-t}{3}}, (1+t)\me^{-t}]}$]{
        \resizebox{7cm}{3cm}{\includegraphics{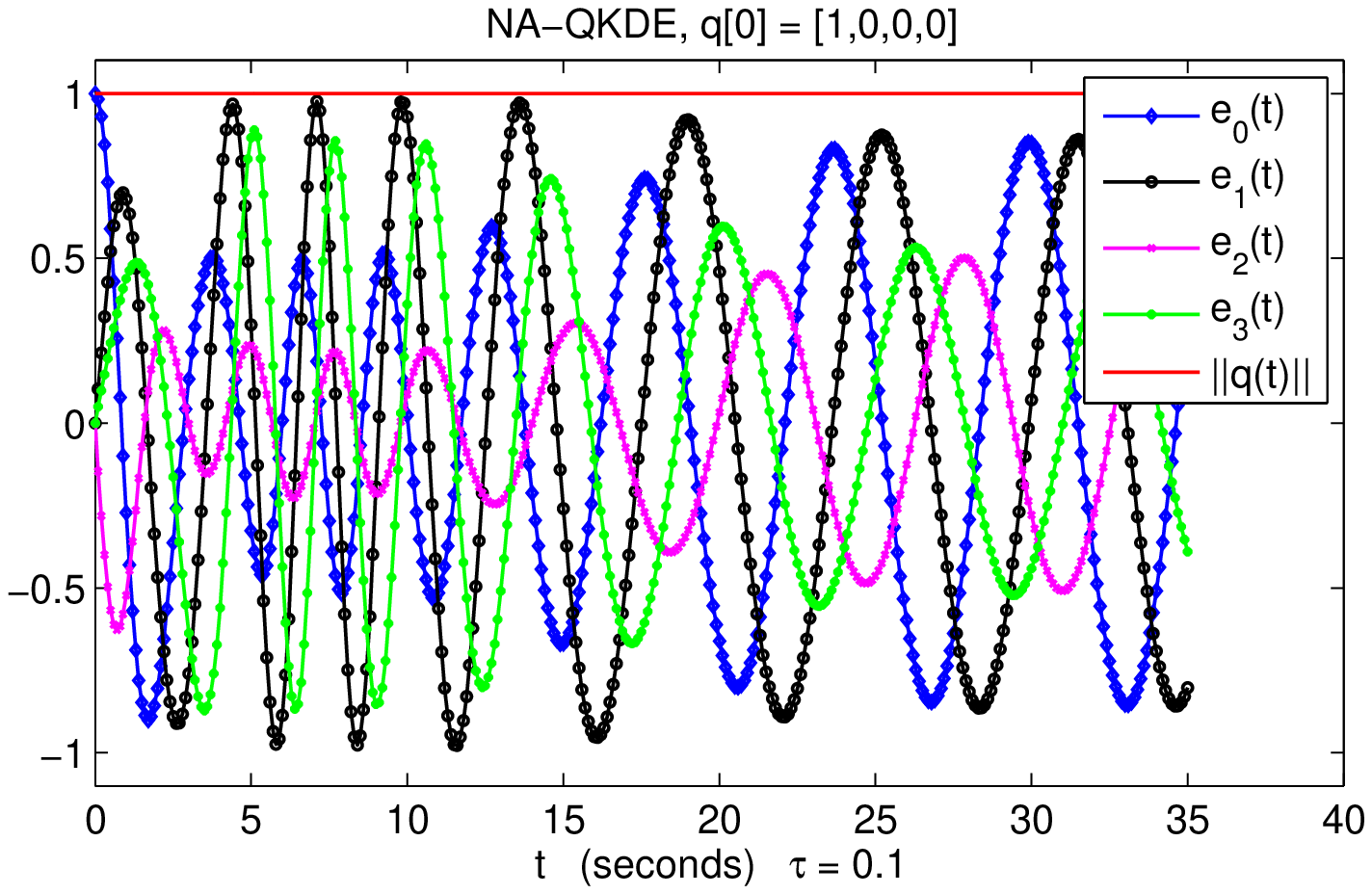}}
        \label{fig-1-Case-3}
  }
  \subfigure[$\vec{\omega}=\trsp{[\sin(10t)-2, 2 t + 1.4, 4 - 0.2\cos(3t)]}$]{
        \resizebox{7cm}{3cm}{\includegraphics{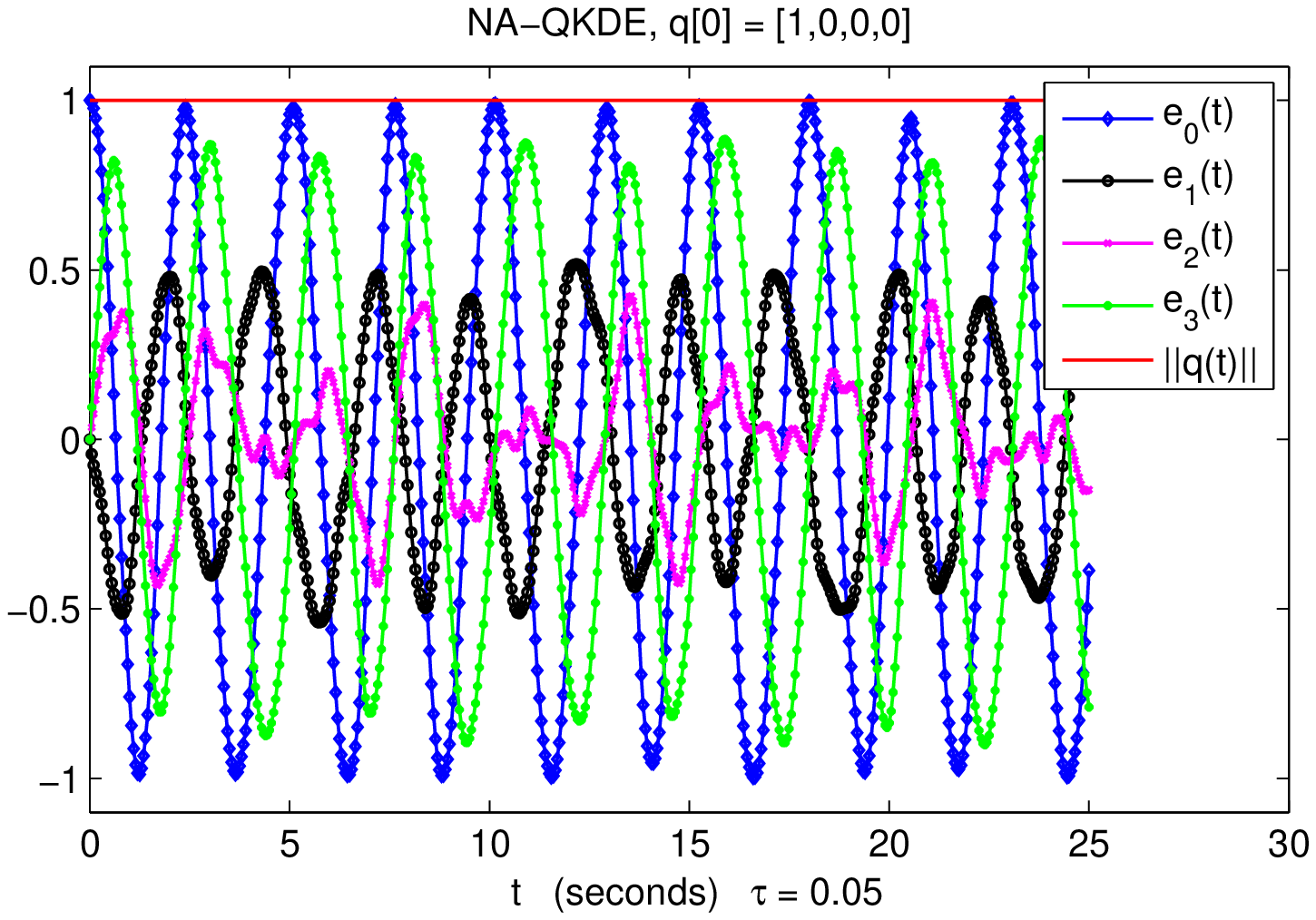}}
        \label{fig-1-Case-4}
  }
\caption{Solution to QKDE with initial state $\quat{q}[0]=\trsp{[1,0,0,0]}$ with SAG for different angular velocities.}
\label{fig-1-omega}
\end{figure}

Fig. \ref{fig-1-omega} demonstrates the asymptotic performance of SAG for QKDE. We find that for different $\vec{\omega}(t)$, the norm $\norm{\quat{q}(t)}=1$ is kept and there are no accumulated errors. In Fig.\ref{fig-1-Case-1}, each $e_i(t)$ is a cosine function  since $\omega_i$ is a constant according to Eq.(\ref{eq-ei-cosine}). In Fig.\ref{fig-1-Case-2}, $e_0(t)$ and $e_1(t)$ behave like cosine curves asymptotically because $\omega_1(t)\to 0$ asymptotically. Moreover, it is trivial that $e_2(t) = e_3(t) \equiv 0$ since $\omega_2(t) = \omega_3(t) \equiv 0$. In Fig.\ref{fig-1-Case-3}, each  $e_i(t)$ varies periodically when $t>10$ by Eq.(\ref{eq-38}).
Although the each $e_k(t)$ varies
independently, the norm $\quat{q}(t) \equiv 1$.

\begin{figure}[t]
  \subfigure[large step, $\tau = 0.25$]{
        \resizebox{7cm}{5cm}{\includegraphics{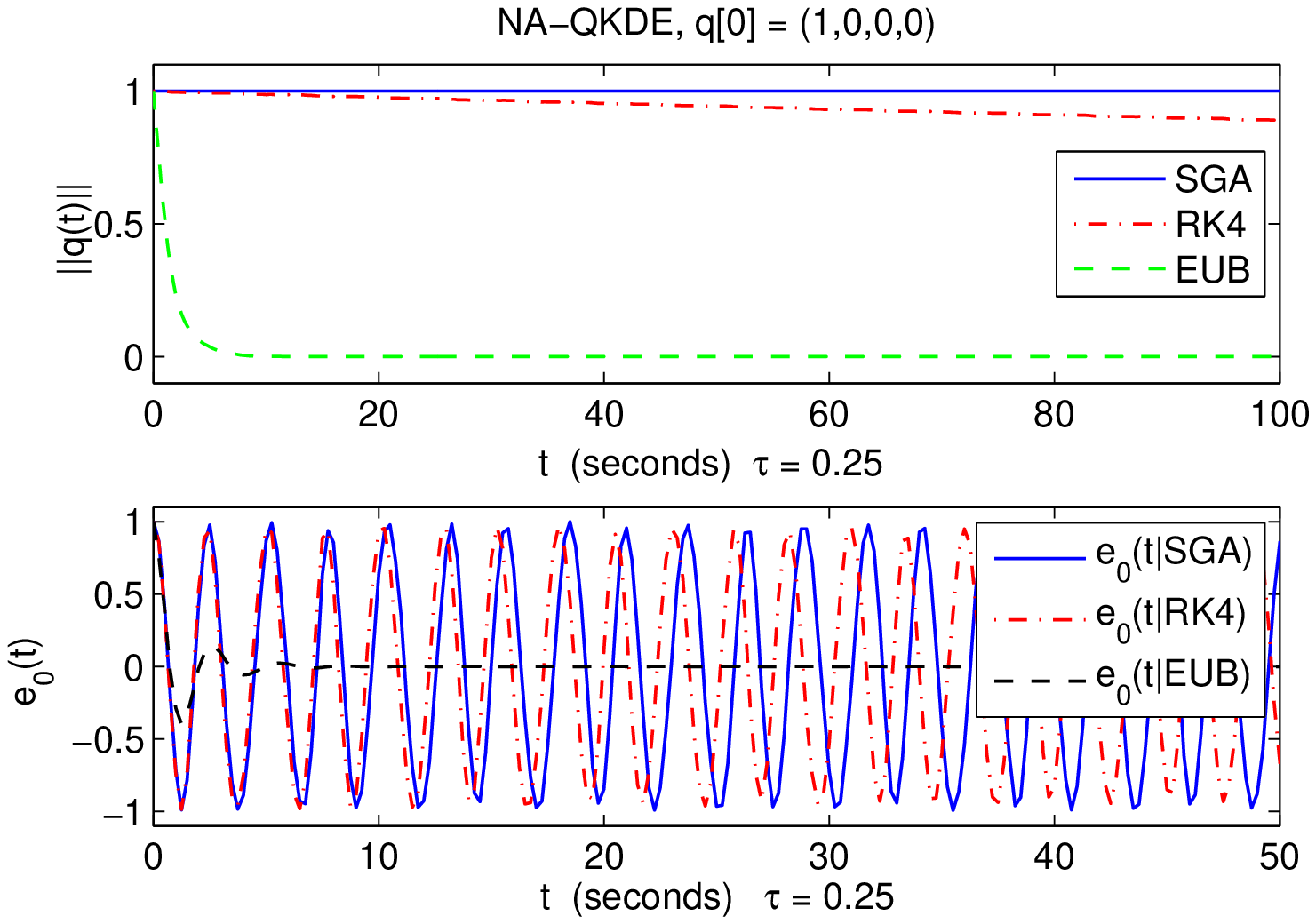}}
		\label{fig-SGA-RK4-EUB-1}
   }
  \subfigure[small step, $\tau = 0.10$]{
        \resizebox{7cm}{5cm}{\includegraphics{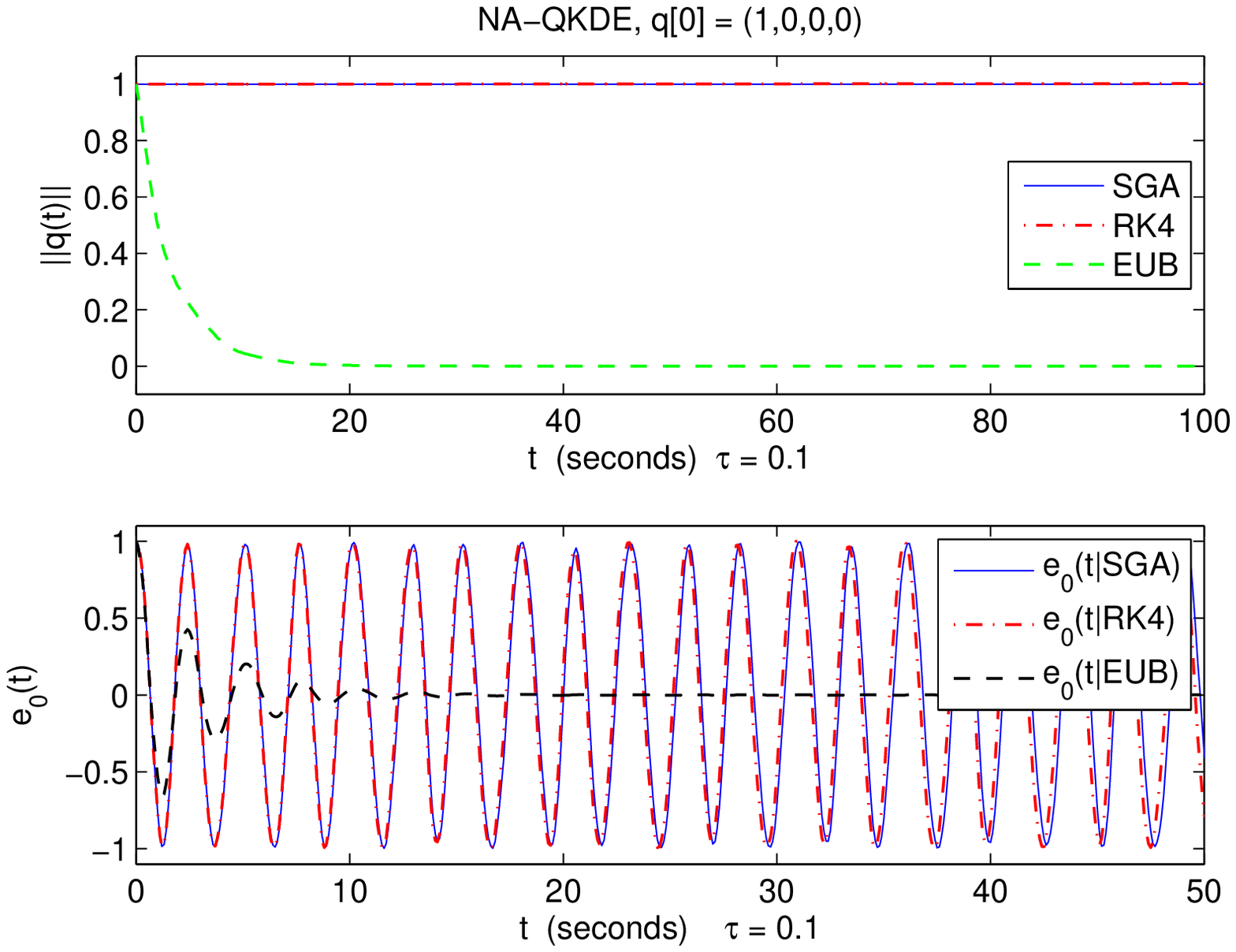}}
        \label{fig-SGA-RK4-EUB-2}
  }
  \caption{Performances of SGA, RK4 and EUB for QKDE in the case with $\vec{\omega}=\trsp{[\sin(10t) -2,2\sin(t) + 1.4,4-0.2\cos(3t)]}$.
  }
  \label{fig-2-DiffScheme}
\end{figure}

Fig.\ref{fig-2-DiffScheme} shows the precision and stability with the time step. We find that the four-stage explicit Runge-Kutta (RK4) method works well only when the time step is small in Fig.\ref{fig-SGA-RK4-EUB-2} and the time duration is relatively short (15 seconds) in Fig.\ref{fig-SGA-RK4-EUB-1}, and the Euler-Backward (EUB) method always leads to serious accumative errors. However, the SGA works well  and there is no computational damp since the $\norm{\quat{q}} $ remains constant.

\begin{figure*}[!t]
  \centering
  \includegraphics[width=15cm,clip,height= 8cm]{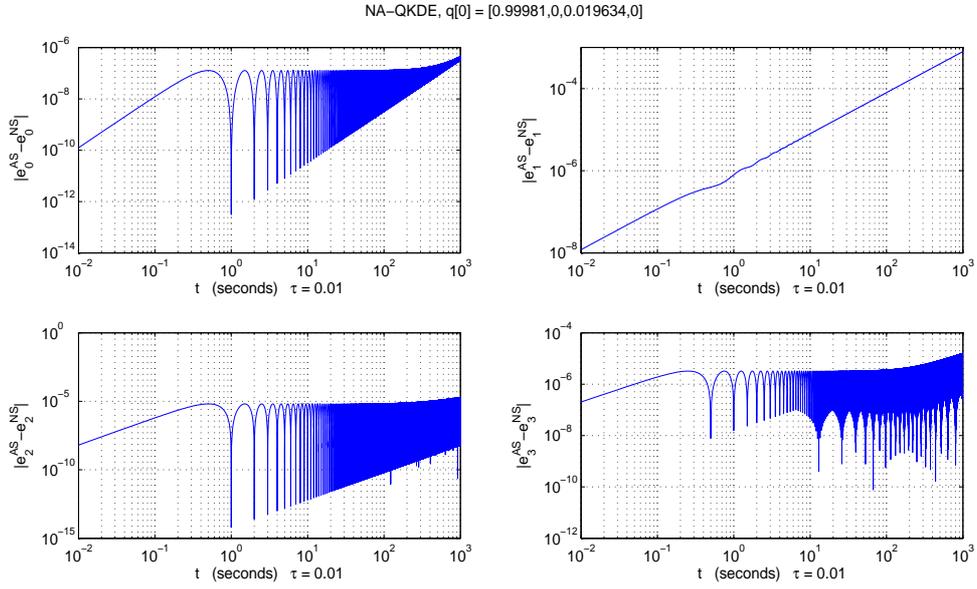}
  \caption{Absolute errors of numerical solutions to NA-QKDE($\omega_0 = 2\pi,\beta = \pi/80$) by using Algorithm \ref{Sp-NA-QKDE} with $\tau=0.01$s, $[t_0,t_f]=[0,1000]$.}\label{fig-NA-QKDE-longTime1}
\end{figure*}

Fig.\ref{fig-NA-QKDE-longTime1} illustrates the performace of SAG for NA-QKDE. We set parameters $\omega_0 = 2\pi, \beta =\pi/80$, initial state $\quat{q}[0]=[\cos (\pi/160),0,\sin(\pi/160),0]$ and
$$ \vec{\omega}(t) =\trsp{[
-\omega_0(1-\cos \beta), -\omega_0 \sin \beta \sin (\omega_0 t),\omega_0 \sin \beta \cos (\omega_0 t)]}
$$
such that the AS~
$$\quat{q}(t) = \trsp{[e_0(t), e_1(t), e_2(t), e_3(t)]}
 =\trsp{\left[\cos(\beta/2), 0, \sin (\beta/2) \cos(\omega_0 t),
\sin(\beta/2)\sin(\omega_0 t)\right]}.$$  We have computed the absolute differences between the numerical solutions and the analytical solutions with Algorithm \ref{Sp-NA-QKDE} for every component of quaternions $|e_i^{AS}-e_i^{NS}|$  $(i\in \{0,1,2,3\})$, as it is shown in  Fig.\ref{fig-NA-QKDE-longTime1}, where the time duration is 1000 s ($[t_0, t_f] = [0, 1000]$). We find that all these
errors of numerical solutions are stably held every small during 1000 s, and the magnitude of error of the quaternion $e_0$ is $10^{-7}$. The accuracy and stability of SGA are proved well for NA-QKDE on long time interval.

\begin{figure}[t]
  \subfigure[maximum errors and computing time with the time step]{
        \resizebox{7cm}{6.2cm}{\includegraphics{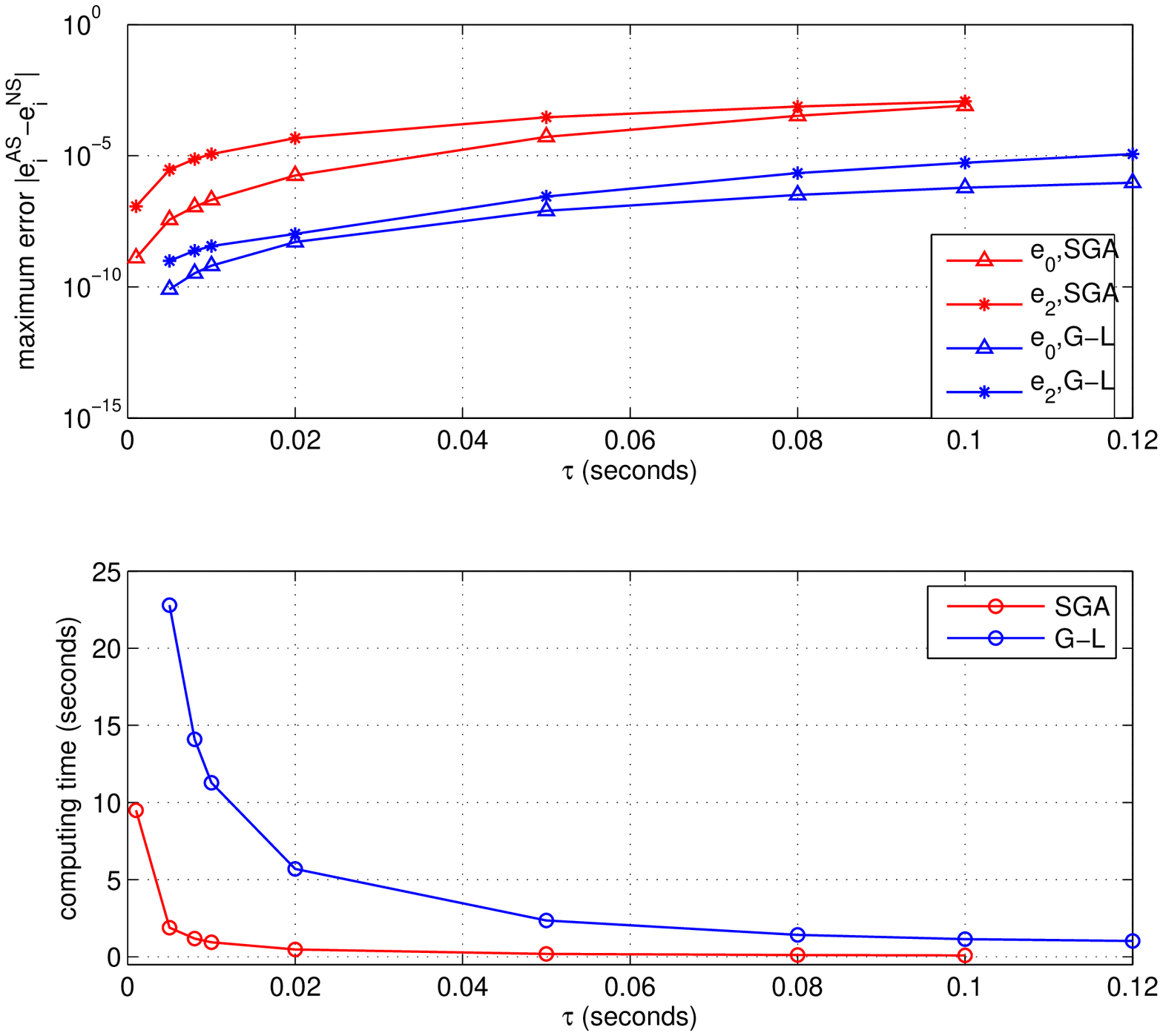}}
		\label{fig-NA-QKDE-longTime2a}
   }
  \subfigure[maximum errors with the computing time]{
        \resizebox{7cm}{6.2cm}{\includegraphics{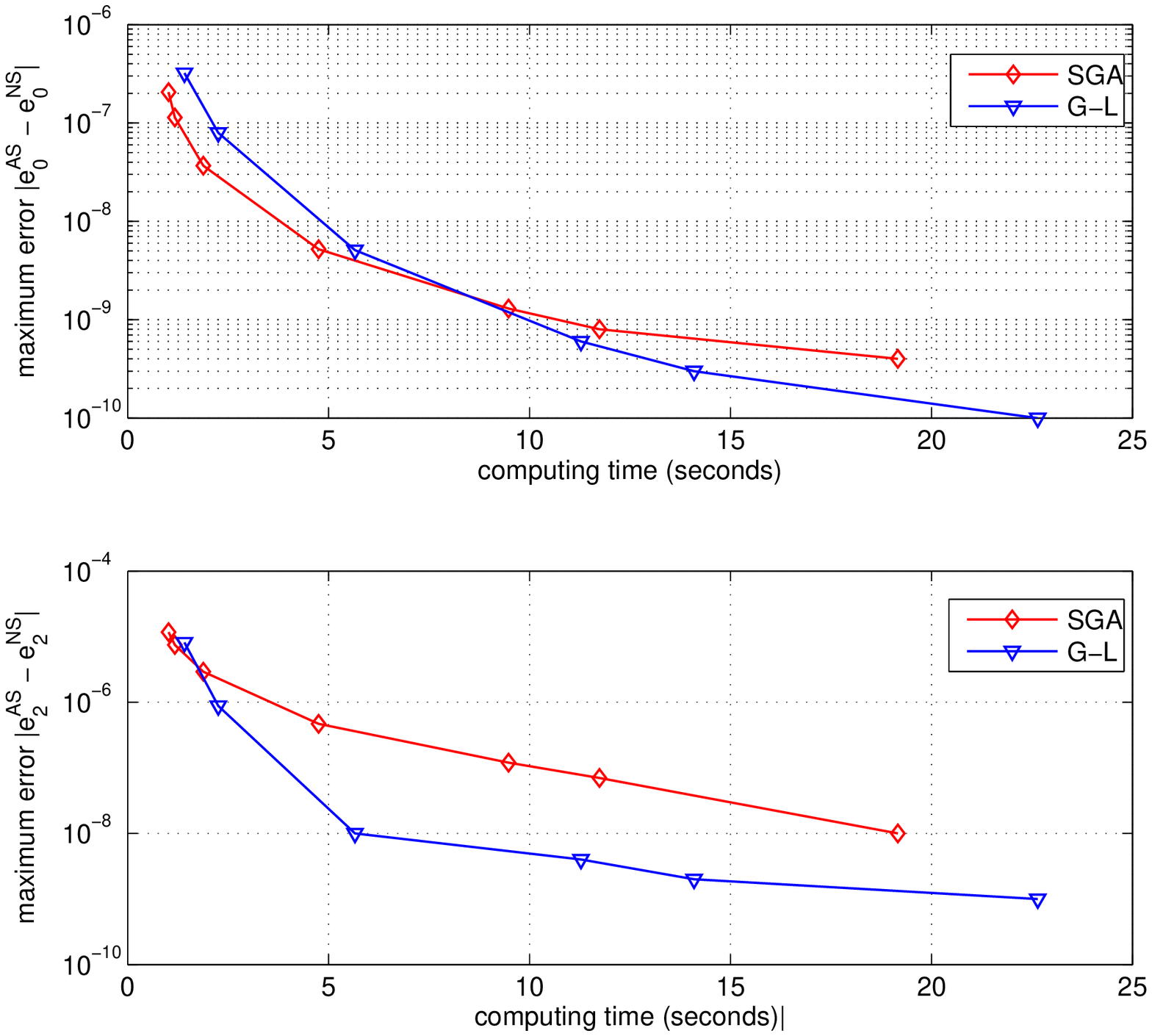}}
        \label{fig-NA-QKDE-longTime2b}
  }
  \caption{Comparison of the performances of SGA and G-L for NA-QKDE($\omega_0 = 2\pi,\beta = \pi/80$) within 500 s.
  }
  \label{fig-NA-QKDE-longTime2}
\end{figure}

Fig.\ref{fig-NA-QKDE-longTime2} represents the low time complexity of SGA for NA-QKDE in comparison with the Gauss-Legendre (G-L) method. G-L method is an accurate and implicit method for the differential equations of time-varying system, however, it needs a lot of computing time to solve the equations. We solved the Eq.(\ref{eq-QKDE})
with the same parameters $\omega(t),~\quat{q}(0)$ as Fig.\ref{fig-NA-QKDE-longTime1} during 500 s, and calculated the maximum error and computing time of the two methods with a series of time step $\tau$.
The numerical results of
quaternion $e_0(t)=\cos(\beta/2)$ and $e_2(t)=\sin(\beta/2)\cos(\omega_0 t)$
were plotted in Fig.\ref{fig-NA-QKDE-longTime2}. Although the precision of our proposed algorithm is worse than G-L in
Fig.\ref{fig-NA-QKDE-longTime2a}, we
point out the time complexity of SGA is far lower than G-L and the computing time of SGA is almost less than 1 s for most cases.
Low time complexity is essential for the real-time system.
In addition, we find that the maximum error
of $e_0$ calculated by SGA is less than by G-L when the computing time
is limited within 7 s in Fig.\ref{fig-NA-QKDE-longTime2b}, from which we can see that
our algorithm performs more competitive in the case
with the limitation of computing time.

\section{Conclusions}\label{section6}

In this paper we proposed the key idea of solving the QKDE with symplectic method: each QKDE can be described by a corresponding Hamiltonian system and the CEIS-2 or T-CEIS-4 can be used to design SGA for QKDE. The Euler's formula for the symplectic transition mapping plays a key role in designing SAG for A-QKDE with first-order precision and NA-QKDE with second-order precision, which simplifies the original algorithmic structures and guarantees the low time complexity of computation.
The correctness and efficiencies of the SAG presented are verified and demonstrated by comparison with AS and NS. Examples show that our proposed algorithms perform very well since the norms of the quaternions can be preserved and the errors accumulated
can be eliminated.

As part of future work, we will investigate the high-order precision SGA \cite{Iwatsu2009} to QKDE, the robustness of the SGA to QKDE perturbed by noise, the applications of SGA to more general linear time-varying system and its combination with precise-integration method \cite{Deng2001Precise}. Additionally, we will also design second-order precision symplectic-precise integrator to solve the linear quadratic regulator problem and the matrix Ricatti equation since they  play an important role in automation.

\section*{Acknowledgments}

The financial support of the Fundamental Research Funds for the Central Universities of China under grant number ZXH2012H005 is gratefully acknowledged. This work was also supported in part by  the National Natural Science Foundation of China (NNFSC) under grant numbers NNSFC  61201085, NNSFC 51205397 and NNSFC 51402356.
The first author would like to thank Prof. Daniel Dalehaye of  Ecole Nationale de l'Aviation Civile (ENAC) because part of this work was carried out while this author was visiting ENAC. We are indebted to Prof. Yifa Tang whose   lectures on symplectic geometric algorithms for Hamiltonian system in Chinese Academy of Sciences attracted the first author and stimulated this work in some sense. The authors' thanks go to Ms. Fei Liu, Mr. Jingtang Hao and Ms. Yajuan Zhang for correcting some errors in an earlier version of this paper.

\end{document}